\documentclass[twocolumn,showpacs,amsmath,amssymb]{revtex4}
\usepackage[dvips]{graphics}
\def\imo{i}
\def\Re#1{Re(#1)}
\def\Im#1{Im(#1)}
\begin{document}
\title{Quasinormal modes, scattering and Hawking radiation of Kerr-Newman black holes in a magnetic field}
\author{K. D. Kokkotas, R. A. Konoplya}\email{konoplya_roma@yahoo.com}
\affiliation{Theoretical Astrophysics, Eberhard-Karls University of T\"{u}bingen, T\"{u}bingen 72076, Germany}
\author{A. Zhidenko}\email{olexandr.zhydenko@ufabc.edu.br}
\affiliation{Centro de Matem\'atica, Computa\c{c}\~ao e Cogni\c{c}\~ao, Universidade Federal do ABC,\\ Rua Santa Ad\'elia, 166, 09210-170, Santo Andr\'e, SP, Brazil}

\begin{abstract}
We perform a comprehensive analysis of the spectrum of proper oscillations (quasinormal modes), transmission/reflection coefficients and Hawking radiation for a massive charged scalar field in the background of the Kerr-Newman black hole immersed in an asymptotically homogeneous magnetic field. There are two main effects: the Zeeman shift of the particle energy in the magnetic field and the difference of values of an electromagnetic  potential between the horizon and infinity, i.e. the Faraday induction. We have shown that ``turning on'' the magnetic field induces a stronger energy-emission rate and leads to ``recharging'' of the black hole. Thus, a black hole immersed in a magnetic field evaporates much quicker, achieving thereby an extremal state in a shorter period of time. Quasinormal modes are moderately affected by the presence of a magnetic field which is assumed to be relatively small compared to the gravitational field of the black hole.
\end{abstract}

\maketitle

\section{Introduction}

Magnetic field is one of the most important constituents of the cosmic space and one of the main sources of the dynamics of interacting matter in the Universe. Weak magnetic fields of about a few $\mu G$ exist in galaxies and clusters of galaxies, while very strong magnetic fields of up to $10^{4}-10^{8} G$ are supposed to exit near supermassive black holes in the active galactic nuclei and even around stellar mass black holes \cite{Han:2006nx,Zhang:2005yq,Piotrovich:2010aq}. Magnetic field near a black hole leads to a number of processes, such as extraction of rotational energy from a black hole, known as the Blandford-Znajek effect \cite{Blandford:1977ds}, the charging of a black hole due to accretion of charged matter \cite{Aliev:1989ww}, the formation of an induced electric field on the black hole surface \cite{Aliev:1989wx}, negative absorption (masers) of electrons \cite{Aliev:1986wu}, and so on. In addition to stellar mass and galactic black holes, miniature black holes could be immersed in a strong magnetic field if created in a laboratory or observed in cosmic showers \cite{Kanti-review}.

Even a relatively weak magnetic field can considerably affect the behavior of charged particles/fields due to usually not weak coupling $e B$ between the particle charge $e$ and the magnetic field $B$. Therefore, charged massive fields are interesting models  for theoretical study of interaction of a magnetized black hole with its surroundings. As the simplest case, one may neglect the spin of the field and consider the complex massive charged scalar field. Still, the interaction of particles due to the spin can also significantly affect the particles' state and thus deserves a separate consideration. As a first step in this direction, we shall consider here a charged rotating black hole, given by the Kerr-Newman solution, and a charged massive scalar field propagating on its background and immersed in an asymptotically homogeneous magnetic field. The magnetic field is supposed to be weak enough so that the metric does not deviate from the Kerr-Newman one, i.e. the magnetic field does not distort the geometry of the space-time  but only interacts with other electromagnetic charges in the system.

Particles and fields in the vicinity of a black hole slightly change the background space-time of a system. Therefore, the addition of a field to a black hole space-time can be considered as a perturbation. At the classical level, the perturbation can be described by its damped characteristic modes, called the quasinormal modes (QNMs) \cite{Kokkotas:1999bd,Nollert:1999ji} and by the scattering properties, which are encoded in the $S$-matrix of the perturbation. Quasinormal modes are proper oscillations of the perturbation which dominate at late time in the response of a black hole to z perturbation. The complex frequencies of such oscillations do not depend on the manner of excitation but only on the parameters of the black hole and the field under consideration. Therefore, they are usually called the ``fingerprints'' of a black hole.

In the same way, as quasinormal modes are an essential classical characteristic of a black hole, the thermal Hawking radiation is its essential quantum feature that carries information about the dynamics of evaporation of the black hole. For large astrophysical black holes, the effect of Hawking evaporation is certainly negligible for the black hole dynamics but not for the behavior of particles in its vicinity. Emission of Hawking radiation is significant  for primordial black holes and huge for miniature black holes which are considered in the higher dimensional gravity and string theory. According to brane-world scenarios, our world is assumed to be a (3+1)-dimensional brane which is embedded in a higher dimensional bulk. Gravity is supposed to be much stronger at small distance creating possibilities for the formation of an event horizon in particle collisions even at energies achievable at the Large Hadron Collider \cite{Kanti-review}. Estimates show that, once such a black hole is created, it will almost immediately evaporate, so that the life-time of the miniature black holes is about $10^{-25} -10^{-32}$ sec. Although created in this way miniature black holes would be intrinsically higher dimensional. Nevertheless, our consideration here of the 4-dimensional black holes can be useful for two reasons: First, it is known that the higher dimensional black hole emits particles mainly ``on the brane,'' i.e. the process of quantum evaporation of the higher dimensional black holes is probably qualitatively similar in many aspects to the one for 4-dimensional black holes. Second and more important, in the proposed approach, we can realize how a magnetic field can influence the process of evaporation.  Thus, when talking about QNMs and Hawking radiation of black holes immersed in a magnetic field we have in mind not only astrophysical black holes but also primordial and miniature black holes.

Thus, we could say that the quasinormal spectrum and Hawking radiation are, respectively, classical and quantum ``fingerprints'' of a black hole.
Quasinormal modes and Hawking radiation also have one technical point in common: analysis of QNMs as well as of the Hawking radiation (in semiclassical approximation) begins from the linear perturbations of the fields under consideration whose dynamics should be reduced to a single wavelike equation, called the master equation.

Up to now there are two kinds of analysis of quasinormal modes which, in a sense, are complimentary to the present work. Quasinormal modes of the massive and massless charged scalar field around charged black holes (without a magnetic field) were studied in \cite{Konoplya:2002wt,Konoplya:2002ky,Konoplya:2007zx}. Quasinormal modes of a neutral scalar field around black holes immersed in a strong magnetic field were calculated in \cite{Konoplya:2007yy,Konoplya:2008hj}. In \cite{Konoplya:2007yy,Konoplya:2008hj}, the black hole was described by the Ernst-Schwarzschild solution which contains a magnetic field as a parameter because the magnetic field is implied to be strong enough in order to deform the black hole geometry significantly. However, such strong geometry-deforming  magnetic fields have little probability of existing in nature \cite{Piotrovich:2010aq}.

Here, we shall consider a more realistic situation and assume that the magnetic field is not strong enough to deform the Kerr-Newman black hole metric. The correlation of the quasinormal frequencies, the reflection coefficients and the energy emission rates with the parameters of the black hole (mass $M$, charge $Q$, angular momentum $a$) and of the scalar field (mass $\mu$, charge $e$) are analyzed here through a comprehensive numerical study.
The Hawking radiation for charged particles (without a magnetic field) around nonrotating and rotating black holes was considered
in \cite{Page:1977um,Page:1976jj} for 4-dimensional black holes and in \cite{Sampaio:2009tp,Sampaio:2009ra} for higher
dimensional scenarios.

In the system under consideration, the coupling of particle charge $e$ with the magnetic field $B$ leads to the Zeeman shift of the energy  $\mu^2 \rightarrow \mu^2 - e B m$ \cite{Galtsov-book}. The rotation of the black hole in the magnetic field, in its turn, leads to the appearance of the induced charge on the black hole surface and to the difference in values of the electromagnetic potential at the horizon and at infinity, that is, the Faraday induction. We shall observe how these two effects, the Zeeman effect and the Faraday induction, are reflected in the processes of classical and quantum radiation. Qualitatively, these two effects were considered in the vicinity of the Kerr-Newman black hole by Galtsov and collaborators \cite{Aliev:1989wx}. Here we shall give an accurate quantitative analysis for the above case. We shall calculate characteristic quasinormal modes, reflection/transmission coefficients, and the emission rates for Hawking radiation of the charged massive scalar field in the background of the Kerr-Newman black holes and in the vicinity of the asymptotically homogeneous magnetic field.

The paper is organized as follows. Section II is devoted to the separation of variables for the scalar field in the Kerr-Newman background under nonzero magnetic field. In Section III we describe the numerical procedures for finding eigenvalues of the separated angular equation and quasinormal frequencies and reflection/transmission coefficients. Section IV is devoted to calculations of the quasinormal modes. Section V discusses classical scattering and calculates the energy and momentum emission rates for the Hawking radiation. We present our conclusions in Sec. VI.

\begin{widetext}
\section{Wave-like equation}

The Kerr-Newman metric can be written in the following form:
\begin{equation}
ds^2 = \frac{\triangle}{\Sigma} (dt -a \sin^2 \theta d \phi)^{2} - \frac{\sin^2 \theta}{\Sigma} [a d t- (r^2 + a^2)d\phi]^{2} - \frac{\Sigma}{\triangle} dr^2 -\Sigma d \theta^2,
\end{equation}
where
\begin{equation}
\triangle = r^2 - 2 M r +a^2 +Q^2, \quad \Sigma = r^2 +a^2 \cos^2 \theta.
\end{equation}
Here $M$ is the black hole mass, $Q$ is its charge, and $a$ is the angular momentum per unit mass.
The event horizons are situated at
\begin{equation}
r= r_{\pm} = M \pm \sqrt{M^2- a^2 - Q^2}.
\end{equation}

In the above description we have not taken into account the influence of the magnetic field onto the black hole background. Under these conditions the background electromagnetic field can be written as
\begin{equation}
A = A_{\mu} dx^{\mu} = \frac{Q r}{\Sigma} (dt - a \sin^2 \theta d\phi).
\end{equation}
The KN metric does not depend on the coordinates $t$ and $\phi$, so that there exist the two Killing vectors $\xi_{(t)} = (1, 0, 0, 0)$ and $\xi_{(\phi)} =(0, 0, 0, 1)$. One can see that the Killing vectors for vacuum metrics satisfy the same equations as the 4-potentials $A_{\mu}$. This suggests the following form of the 4-potential:
\begin{equation}
A^{\mu} = \frac{1}{2} B \left[\xi_{(\phi)}^{\mu} + 2 a \xi_{(t)}^{\mu} \right] -\frac{Q}{2 M} \xi_{(t)}^{\mu}.
\end{equation}
The gauge transformations,
\begin{equation}
A_{\mu} \rightarrow A_{\mu} + \frac{\partial}{\partial x^{\mu}} \left(\left(\frac{Q- 2 a M B}{2 M} \right) t \right),
\end{equation}
reduce the 4-potential to the Coulomb form $A_{0}(r=\infty) = 0$.
Thus, the ``full'' electromagnetic 4-potential of the system which includes the electric field of the
charged black hole as source and a magnetic field $B$ living ``in the background'' of a Kerr-Newman black hole
has the form
$$A^{\mu}=\left(\frac{(Q-2aMB)r(r^2+a^2)}{\triangle\Sigma},0,0,\frac{B}{2}+\frac{(Q-2aMB)ra}{\triangle\Sigma}\right)$$
\begin{equation}
A_{\mu} A^{\mu} = -\frac{B^2 \sin^2 \theta}{4 \Sigma}  \left[(r^2 + a^2)^{2} - \triangle a^{2} \sin^{2} \theta \right] - (Q - 2 a M B) \frac{a B}{\Sigma}
r \sin^2 \theta + \frac{(Q - 2 a M B)^2 r^2}{\triangle \Sigma}.
\end{equation}
The Klein-Gordon equation for the charged massive scalar field in the vicinity of a Kerr-Newman black hole and in the presence of the homogenous magnetic field $B$ has the following general covariant form \cite{Galtsov-book}:
\begin{equation}
g^{\mu \nu} (\nabla_{\mu} + i e A_{\mu}) (\nabla_{\nu} + i e A_{\nu}) \Psi + \mu^2 \Psi = 0
\end{equation}
Since $\nabla_\mu A^\mu=0$, the latter can be reduced to the following form \cite{Galtsov-book}:
$$\frac{\partial}{\partial r} \left(\triangle \frac{\partial \Psi}{\partial r} \right) +
\frac{1}{\sin \theta} \frac{\partial}{\partial \theta} \left(\sin \theta \frac{\partial \Psi}{\partial \theta} \right) -
\left(\frac{(r^2 + a^2)^2}{\triangle} - a^2 \sin^2 \theta \right) \frac{\partial^{2}\Psi}{\partial t^2} +2a \left(1-\frac{r^2+a^2}{\triangle}\right)
\frac{\partial^{2}\Psi}{\partial t \partial \phi} + \left(\frac{1}{\sin^2 \theta} - \frac{a^2}{\triangle}\right) \frac{\partial^2 \Psi}{\partial^2 \phi}$$
\begin{equation}
- 2 i e \left[ \frac{r(r^2 + a^2) (Q - 2 a M B)}{\triangle} \frac{\partial \Psi}{\partial t} + \left( \frac{(Q - 2 a M B)ra}{\triangle
} + \frac{B\Sigma}{2} \right) \frac{\partial \Psi}{\partial \phi} \right] + (e^2 A_{\mu} A^{\mu} - \mu^2) \Sigma \Psi = 0.
\end{equation}

As was shown in \cite{Galtsov-book}, the separation of radial and angular variables in the whole space is impossible for this equation.
Nevertheless, if one considers only the region which begins at the event horizon and ends at some distance far from
the black hole $r \gg r_+$, and uses the following approximations:
\begin{equation}\label{approximation}
e B r_+^2 \ll 1, \quad e Q \ll1,
\end{equation}
then we get,
$$A_{\mu} A^{\mu}=\frac{(Q - 2 a M B)^2 r^2}{\triangle \Sigma}.$$
Under these conditions the separation of variables is possible, that is, we can write
\begin{equation}\label{anzats}
\Psi = e^{-i \omega t + i m \phi} S(\theta) R(r)/\sqrt{r^2+a^2}.
\end{equation}
Here the function $S(\theta)$ obeys the equation
\begin{equation}\label{angularpart}
\left(\frac{\partial^2}{\partial \theta^2} + \cot \theta \frac{\partial}{\partial\theta} - \frac{m^2}{\sin^2 \theta} - (a \omega)^2\sin^2 \theta + 2 m a \omega + \lambda - (\mu^2 - e B m) a^2 \cos^2 \theta \right) S(\theta) = 0.
\end{equation}
This equation can be solved numerically for any value of $\omega$ in the same way as the equation for massive scalar field in the Kerr black hole background \cite{Konoplya:2006br} with the effective mass $\mu_{\rm eff}^2=\mu^2 - e B m$. We note that, when $\mu_{\rm eff}=0$, Eq. (\ref{angularpart}) reduced to the well-known equation for the spheroidal functions. In this case, the separation constant $\lambda(\omega)$ can be found numerically using the continued fraction method \cite{Suzuki:1998vy}. When the effective mass is not zero, the separation constant can be expressed, in terms of the separation constant for spheroidal functions, as
$$\lambda(\omega,\mu_{\rm eff})=\lambda(\Omega)+2ma(\Omega-\omega)+\mu_{\rm eff}^2a^2,$$
where
$$\Omega=\sqrt{\omega^2-\mu_{\rm eff}^2}=\sqrt{\omega^2-\mu^2+eBm}.$$

The sign of $\Omega$ might be chosen here arbitrarily, but we fix it so that $Re(\Omega)$ and $Re(\omega)$ are of the same sign. This allows one to recover easily the limit of $\mu_{\rm eff}=0$ and, later, simplify fixing of the boundary conditions for the radial part.
When $a=0$, one can find that $\lambda=\ell(\ell+1),~\ell=0,1,2\ldots$. For the nonzero values of $a$, the separation constant can be enumerated by the integer multipole number $\ell\geq|m|$.

Using the new tortoise coordinate $r_{*}$, the radial part can be written as a wavelike equation,
\begin{equation}\label{radialpart}
\left(\frac{d^2}{d r_*^2} - V(r)\right) R(r) = 0,
\end{equation}
where  $$ r_*=\frac{(r^2+a^2)}{\triangle}dr,$$
and where the effective potential has the following form
\begin{equation}\label{Effective-potential}
V(r)=\frac{\triangle}{(r^2+a^2)^2}\left(\lambda + (\mu^2 - e B m) r^2+\frac{(r \triangle)'}{r^2 + a^2}-\frac{3 \triangle r^2}{(r^2 + a^2)^2}\right)-\left(\omega - \frac{m a}{r^2 + a^2} - e r\frac{Q-2aMB}{r^2 + a^2}\right)^2.
\end{equation}
From this form of the effective potential one can realize the two effects: the mass of the field gained an effective term  $\mu^2 - e B m$,
and the black hole charge gained an addition as well, $Q \rightarrow Q - 2 a M B$. The first effect is the well-known \emph{Zeeman effect}, which
is the shift of energy of a charged particle (with a charge $e$) in the magnetic field due to interaction of a magnetic field $B$ with
an azimuthal momentum $m$. In systems which are more symmetric than ours, i.e. with degenerated $m$-states, the Zeeman effect leads to splitting of
the $m$-degeneration, and is well known in quantum mechanics. In the case when the effective potential allows for nondegenerated $m$-states, the Zeeman effect simply corresponds to a shift in the particle's energy.

The second effect is more remarkable. Once a rotating black hole is immersed into a magnetic field, the electrostatic potential between the horizon and infinity acquires a difference due to the presence of a magnetic field, which is
$$ \delta A = A_{\rm hor} - A_{\rm inf} = \frac{Q - 2 a M B}{2 M}.$$
In other words the black hole receives an additional
induced electrostatic force $F_{\rm ind} = 2 a M B/r_{+}^{2}$. This is nothing but the \emph{Faraday induction}.
It should be noted that this effect can be significant even for neutral black holes $Q=0$, and can be applied to large astrophysical black holes which cannot possess large electric charge.

The asymptotics of the effective potential near the horizon and at infinity are
\begin{eqnarray}
V(r) &\rightarrow& -\Omega^2, \quad r \rightarrow \infty,\qquad \Omega = \sqrt{\omega^2 - \mu^2 + eBm},\\
V(r) &\rightarrow& -\tilde\omega^2, \quad r \rightarrow r_+, \qquad \tilde\omega = \omega - \frac{ma+er_+(Q-2aMB)}{a^2+r_+^2}.
\end{eqnarray}
During super-radiance the black hole can be charged until it reaches the ``extremal'' charge $Q= 2 a M B$.

In the following sections we will study the influence of the above-mentioned Zeeman and Faraday effects on the classical (quasinormal) and quantum (Hawking) radiation of black holes. However, before starting the numerical study of the wavelike equation (\ref{radialpart}), let us mention one more constraint related to our analysis. If the electric field $2 a M B/r_{+}$ induced on the horizon is as strong as the Schwinger field $\mu^2/e$, then the electrodynamic
process of particle production is initiated and continues until the maximum value of the charge is reached. This maximum values of the charge is
$Q = 2 a M B$. Indeed, imagine vacuum as consisting of virtual pairs $e^{+} e^{-}$ where electrons and positrons, after transforming to the real ones,  become separated by distance of the order of the Compton wavelength $\lambda = 2 \pi \mu^{-1}$. If the work done by the electric field, which is $ e E \lambda$, is as large as the rest mass of the two particles $2 \mu$, then the virtual pair turns into a real one: $2 \pi \mu^{-1} e E > 2 \mu$ and consequently $E \approx \mu^{2}/e$, where $E$ is the induced electric field. We did not take this Schwinger mechanism into consideration in the present paper. However, in the conclusion, we shall suggest simple arguments showing that the Schwinger mechanism of pair production will enhance the process of ``recharging'' a black hole and force the black hole to evaporate faster.

\end{widetext}

\section{Numerical methods}

In this section we shall briefly discuss the two classical numerical methods (Frobenius and WKB) used for calculations of the quasinormal modes and the shooting method used for calculations of the transmission/reflection coefficients.

\subsection{Quasinormal modes}

In order to calculate quasinormal modes we impose the \emph{quasinormal mode boundary conditions} for the wave Eq. (\ref{radialpart}),
i.e. we require that at the black hole horizon we have only purely ingoing waves,
$$R(r_*\rightarrow-\infty)\propto \exp(-\imo\tilde\omega r_*),$$
while we should have only purely outgoing waves at spatial infinity, i.e.
$$R(r_*\rightarrow\infty)\propto \exp(\imo\Omega r_*).$$
Thus, no waves are coming from the horizon or infinity, which implies that $\omega$ are proper oscillation modes in the black hole response to an ``instantaneous'' perturbation. In other words, when the perturbation decays, the source of the initial perturbation is not acting anymore.

Equation (\ref{radialpart}) has an irregular singularity at spatial infinity and four regular singularities at $r=r_+$, $r=r_-=(Q^2+a^2)/r_+$ and $r=\pm\imo a$. The four singularities appear due to the prefactor $(r^2+a^2)^{-1/2}$ in (\ref{anzats}).
The appropriate Frobenius series is determined as
$$R(r)=\left(\frac{r-r_+}{r-r_-}\right)^{-\displaystyle\imo\tilde\omega/4\pi T_H}e^{\displaystyle\imo\Omega r}(r-r_-)^{\displaystyle\imo\sigma}y(r),$$
where
$$\sigma=\left(\Omega+\frac{\mu^2-eB(m-2a\omega)}{2\Omega}\right)(r_++r_-),$$
and $T_H$ is the Hawking temperature
$$T_H=\frac{\Delta'(r_+)}{4\pi(r_+^2+a^2)}.$$
The function $y(r)$ must be regular at the horizon and spatial infinity and
$$y(r)=\frac{\sqrt{r^2+a^2}}{r-r_-}\sum_{k=0}^{\infty}a_k\left(\frac{r-r_+}{r-r_-}\right)^k.$$
The coefficients $a_k$ satisfy the three-term recurrence relation.
\begin{equation}
\alpha_n a_{n+1} + \beta_n a_n + \gamma_n a_{n-1} = 0, \quad n \geq 0, \qquad \gamma_0 = 0,
\end{equation}
where $\alpha_n$, $\beta_n$, $\gamma_n$ can be found in an analytic form. We do not write these coefficients here because they have quite a cumbersome form.
Notice that the factor $\frac{\sqrt{r^2+a^2}}{r-r_-}$ removes the singularities $r=\pm\imo a$ of $y(r)$. Except for this factor, we would have had a five-terms recurrence relation due to the additional singular points.

By comparing the ratio of the series coefficients
\begin{eqnarray}%
\frac{a_{n+1}}{a_n}&=&\frac{\gamma_{n}}{\alpha_n}\frac{\alpha_{n-1}}{\beta_{n-1}
-\frac{\alpha_{n-2}\gamma_{n-1}}{\beta_{n-2}-\alpha_{n-3}\gamma_{n-2}/\ldots}}-\frac{\beta_n}{\alpha_n},\nonumber\\
\label{ratio}\frac{a_{n+1}}{a_n}&=&-\frac{\gamma_{n+1}}{\beta_{n+1}-\frac{\alpha_{n+1}\gamma_{n+2}}{\beta_{n+2}-\alpha_{n+2}\gamma_{n+3}/\ldots}},
\end{eqnarray}%
we obtain an equation with a convergent \emph{infinite continued fraction} on its right side:
\begin{eqnarray}\label{continued_fraction} \beta_n-\frac{\alpha_{n-1}\gamma_{n}}{\beta_{n-1}
-\frac{\alpha_{n-2}\gamma_{n-1}}{\beta_{n-2}-\alpha_{n-3}\gamma_{n-2}/\ldots}}=\qquad\\\nonumber
\frac{\alpha_n\gamma_{n+1}}{\beta_{n+1}-\frac{\alpha_{n+1}\gamma_{n+2}}{\beta_{n+2}-\alpha_{n+2}\gamma_{n+3}/\ldots}},
\end{eqnarray}%
which can be solved numerically by minimizing the absolute value of the difference between its left and right sides. Equation (\ref{continued_fraction}) has an infinite number of roots, but the most stable root depends on $n$. Generally the larger number $n$ corresponds to the larger imaginary part of the root $\omega$ \cite{Leaver:1985ax}.

Note that the case under consideration allows one to use the Nollert procedure \cite{Nollert}, in order to improve convergence of the infinite continued fraction, which is useful for searching roots with a very large imaginary part.

For an additional check of the accurate numerical results obtained by the convergent Frobenius method, we shall use also the WKB formula of the 6th order beyond the eikonal approximation \cite{WKB,WKBorder}. The formula has the following form:
\begin{equation}\label{WKBformula}
	\frac{\imo V_{0}}{\sqrt{2 V_{0}''}} - \sum_{i=2}^{i=6}
		\Lambda_{i} = n+\frac{1}{2},\qquad n=0,1,2\ldots,
\end{equation}
and the correction terms $\Lambda_{i}$ were obtained in \cite{WKB,WKBorder} and depend on higher derivatives of $V$ at its maximum with respect to the tortoise coordinate $r_\star$, and $n$ labels the overtones. The WKB approach was developed by Schutz and Will \cite{WKB} and extended to the 3rd  \cite{WKB} and 6th \cite{WKBorder} orders. It can be effectively used not only for finding low-lying quasinormal modes (see, for instance, \cite{WKBuse,superradiant} and references therein), but also for calculations of the transmission/reflection coefficients in various problems \cite{WKBuse2}.

\subsection{Reflection coefficients}

For calculations of the emission rates of particles due to Hawking radiation, one needs first to solve the problem of classical scattering in order to obtain the gray-body factors. This implies the posing of classical \emph{scattering boundary conditions}. At the event horizon, this again means imposing the boundary condition which corresponds to a purely ingoing wave, while at spatial infinity ($r\rightarrow\infty$) we have a different condition from the one used for the quasinormal modes,
$$R(r)\simeq Z_{in} \exp(-\imo\Omega r_\star)+Z_{out} \exp(\imo\Omega r_\star),$$
where $Z_{in}$ and $Z_{out}$ are integration constants which correspond to the ingoing and outgoing waves respectively. Thus, we would like to know which portion of particles will be able to pass through the barrier of the effective potential.

Introducing the new function
$$P(r)= R(r) \left(\frac{r-r_+}{r-r_-}\right)^{\displaystyle\imo\tilde\omega/4\pi T_H},$$
and choosing the integration constant as $P(r_+)=1,$ we expand Eq. (\ref{radialpart}) near the event horizon and find $P'(r_+)$, which completely fixes the initial conditions for the numerical integration. Then, we integrate Eq. (\ref{radialpart}) numerically from the event horizon $r_+$ to some distant point $r_f\gg r_+$ and find a fit for the numerical solution far from the black hole in the following form:
\begin{equation}\label{fit}
P(r)=Z_{in} P_{in}(r)+Z_{out} P_{out}(r),
\end{equation}
where the asymptotic expansions for the corresponding functions are found by expanding (\ref{radialpart}) at large $r$ as
\begin{eqnarray}
P_{in}(r)&=&e^{-\imo\Omega r}r^{-\displaystyle\imo\sigma}\left(1+P_{in}^{(1)}r^{-1} + P_{in}^{(2)}r^{-2}+\ldots\right),\nonumber\\
P_{out}(r)&=&e^{\imo\Omega r}r^{\displaystyle\imo\sigma}\left(1+P_{out}^{(1)}r^{-1} + P_{out}^{(2)}r^{-2}+\ldots\right).\nonumber
\end{eqnarray}
The fitting procedure allows us to find the coefficients $Z_{in}$ and $Z_{out}$. In order to check the accuracy of the calculated coefficients, one should increase the internal precision of the numerical integration procedure, the value of $r_f$, and the number of terms in the series expansion for $P_{in}(r)$ and $P_{out}(r)$, making sure that the values of $Z_{in}$ and $Z_{out}$ do not change within desired precision.

If the coefficients $Z_{in}$ and $Z_{out}$ are calculated, one can find the absorbtion probability
\begin{equation}\label{absorbtion}
|{\cal A}_{\ell,m} |^2=1-|Z_{out}/Z_{in}|^2.
\end{equation}
This will be used later for calculations of the emission rates for energy momentum and charge of the black hole. This approach was also used for analysis of Hawking radiation of higher dimensional simply rotating black holes \cite{Kanti:2009sn}, and of Gauss-Bonnet black holes \cite{Konoplya:2010vz}, and it showed an excellent agreement with the analytical approach.

\begin{figure}
\includegraphics*{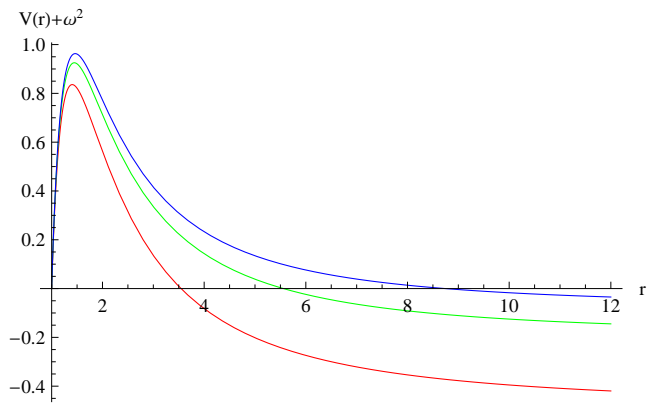}
\caption{The effective potentials for the charged scalar field ($\ell=m=2$) in the Schwarzschild background ($r_+=1$) with the magnetic field $eB=0.04$ (top blue curve), $eB=0.1$ (middle green curve), $eB=0.25$ (bottom red curve). One can see that the potential is negative only deeply in the region $r \gg 1/\sqrt{e B}$, which is beyond the region of validity of the approximation (\ref{approximation}).}
\label{KN-potential-fig}
\end{figure}

\section{Quasinormal modes}

First, let us briefly review previous works on quasinormal modes of Kerr-Newman black holes. Apparently the first work on QNMs of Kerr-Newman black holes was by one of us \cite{Kokkotas:KN}, where gravitational perturbations with a frozen Maxwell field were considered. More accurate numerical results for gravitational perturbations were presented in \cite{Berti:KN}. Quasinormal modes of the Dirac field were considered in \cite{Jing:KN}. The case of charged scalar and Dirac modes was analyzed in the background of Kerr-Newman black holes allowing for a positive cosmological constant \cite{Konoplya:2007zx}. All the above papers observed no unstable modes in the quasinormal spectrum (except the superradiant ones), although unstable modes of a charged scalar field were found for an asymptotically anti-de Sitter background \cite{Elcio-1}.

Before we examine the dependence of QNMs on various parameters of the system, we should first discuss the stability of the system.
The effective potential Eq. (\ref{Effective-potential}) contains a term proportional to $\mu^2 - e B m$ which works as an effective mass term, and, when $\mu^2 < e B m$, this term is negative. It is well known that a massive scalar field with negative $\mu^2$ is unstable even for tiny negative values of the square of mass \cite{Jing-negative-mass}. Thus, the instability is expected when formally considering exact solutions of the wave equations (\ref{radialpart}), but certainly not for a real physical situation. The reason is that the instability due to negative square of mass comes from infinite negative fall-off of the effective potential for scalar field at spatial infinity. In our case, however, ``infinity'' is located at $r \approx r_{+} (e B)^{-1}$, and further
from this distance the wave equation is not valid because of the approximation (\ref{approximation}) which has been used for the separation of variables. In Fig. \ref{KN-potential-fig}, one can see that the effective potential is positive definite in the region of its validity and is negative only for values of $r$ which are seemingly larger than $(e B)^{-1}$.

There are reasons to expect that the true effective potential for the black hole immersed in an asymptotically uniform magnetic field will
inevitably lead to instability due to the infinite energy of the magnetic field. Analysis of particle motion around
Ernst-Schwarzschild and Ernst-Kerr black holes shows that the effective potential for such particles diverges at infinity. This means that the magnetic field which fills in all the Universe will create an effective confining box. Thus, at infinity it will be appropriate to use Dirichlet boundary conditions.
A rotating black hole in such a confining box will inevitably be unstable through the mechanism of superradiance. In a real world the magnetic field is certainly assumed to vanish at infinity, so that no confining box will appear. When using approximation (\ref{approximation}), we ``cut'' the effect of the confining box at infinity in a natural way.

\begin{table}
\caption{QNMs of the massless scalar field in the background of a nonrotating uncharged black hole, $eB=0.05$. The Frobenius method gives the unstable mode, which does not appear when we use the WKB method, supposing an asymptotically flat background.}
\begin{tabular}{|c|c|c|c|}
\hline
mode&unstable&stable&WKB fundamental\\
\hline
$\ell=m=1$&$0.2236\imo$&$0.5747-0.2020\imo$&$0.5747-0.2022\imo$\\
$\ell=m=2$&$0.3162\imo$&$0.9516-0.1988\imo$&$0.9515-0.1989\imo$\\
$\ell=m=3$&$0.3873\imo$&$1.3331-0.1973\imo$&$1.3331-0.1973\imo$\\
$\ell=m=4$&$0.4472\imo$&$1.7162-0.1963\imo$&$1.7162-0.1963\imo$\\
$\ell=m=5$&$0.5000\imo$&$2.0999-0.1956\imo$&$2.0999-0.1956\imo$\\
\hline
\end{tabular}
\end{table}

\begin{figure}
\resizebox{1\linewidth}{!}{\includegraphics*{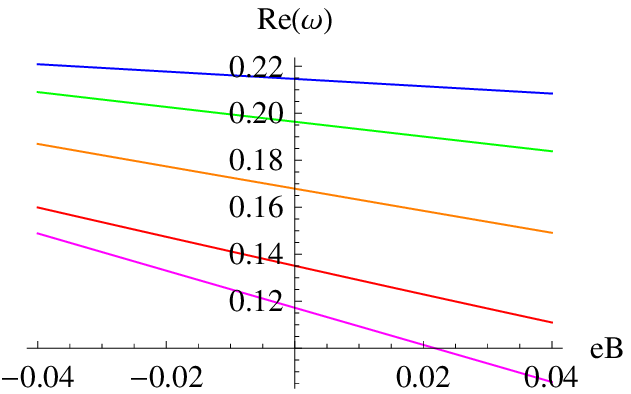}\includegraphics*{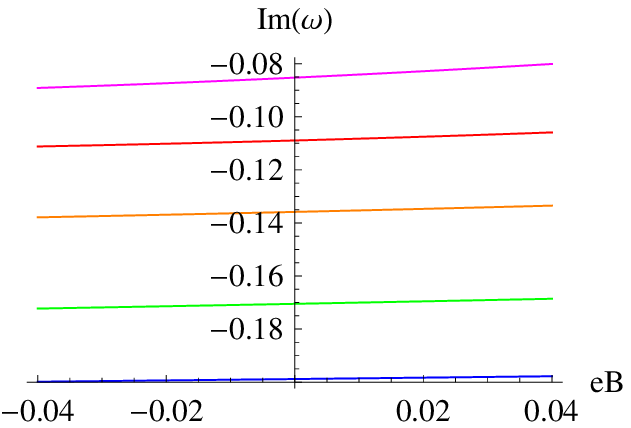}}
\caption{Real (left panel, graphs from top to bottom) and imaginary (right panel, graphs from bottom to top) parts of the fundamental ($n=0$) QNM as a function of $eB$ for $\ell=m=0$, $Q=0$, a=0.2 (blue), a=0.4 (green), a=0.6 (orange), a=0.8 (red), a=0.99 (magenta).}
\label{Fig2}
\end{figure}

\begin{figure}
\resizebox{1\linewidth}{!}{\includegraphics*{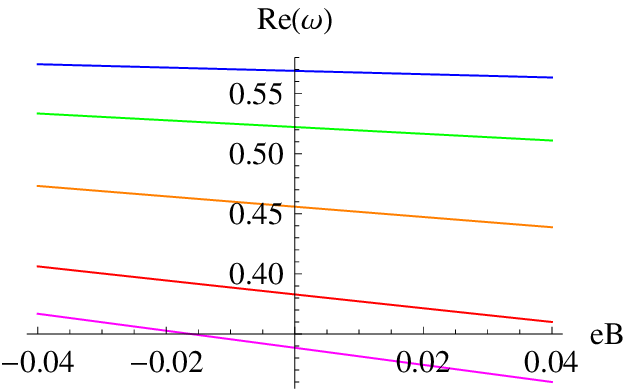}\includegraphics*{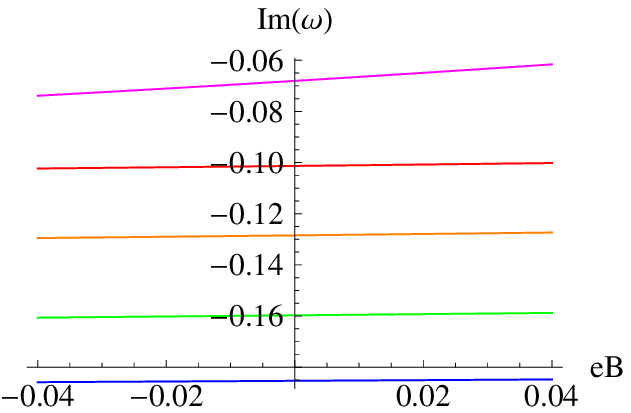}}
\caption{Real (left panel, graphs from top to bottom) and imaginary (right panel, graphs from bottom to top) parts of the fundamental ($n=0$) QNM as a function of $eB$ for $\ell=1$, $m=0$, $Q=0$, a=0.2 (blue), a=0.4 (green), a=0.6 (orange), a=0.8 (red), a=0.99 (magenta).}
\label{Fig3}
\end{figure}

\begin{figure}
\resizebox{1\linewidth}{!}{\includegraphics*{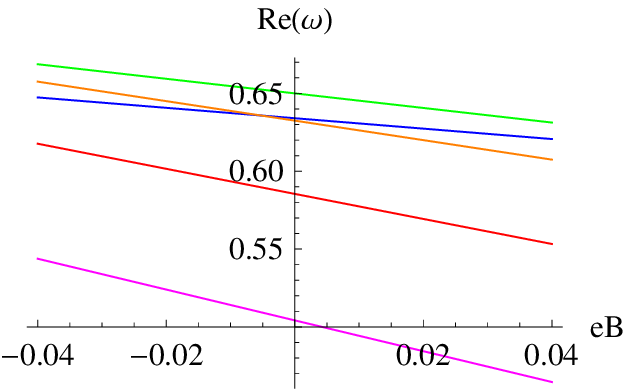}\includegraphics*{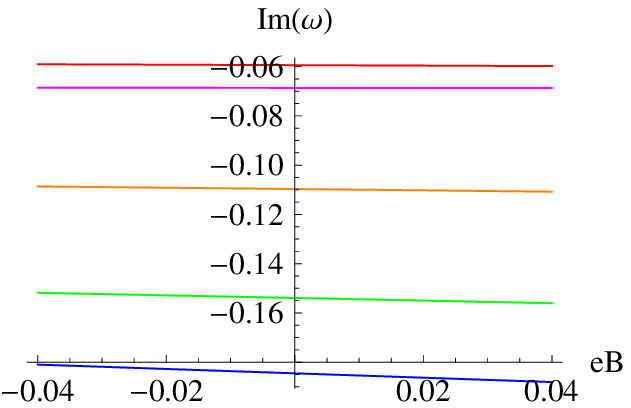}}
\caption{Real (left panel, graphs from top to bottom, except for $a=0.2$) and imaginary (right panel, graphs from bottom to top) parts of the fundamental ($n=0$) QNM as a function of $eB$ for $\ell=1$, $m=1$, $Q=0$, a=0.2 (blue), a=0.4 (green), a=0.6 (orange), a=0.8 (red), a=0.99 (magenta).}
\label{Fig4}
\end{figure}

\begin{figure}
\resizebox{1\linewidth}{!}{\includegraphics*{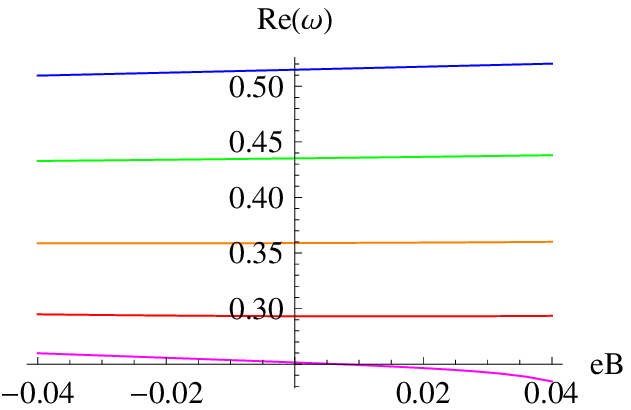}\includegraphics*{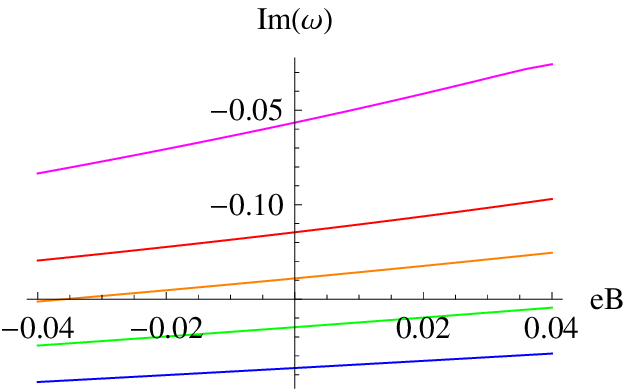}}
\caption{Real (left panel, graphs from top to bottom) and imaginary (right panel, graphs from bottom to top) parts of the fundamental ($n=0$) QNM as a function of $eB$ for $\ell=1$, $m=-1$, $Q=0$, a=0.2 (blue), a=0.4 (green), a=0.6 (orange), a=0.8 (red), a=0.99 (magenta).}
\label{Fig5}
\end{figure}

In nature, infinity means a region far from the black hole $r \gg r_{+}$ which can be approximately treated as asymptotically flat.  In practice, one should match the considered solution with asymptotically homogeneous magnetic field at ``infinity'' $r \approx r_{+} (e B)^{-1}$ with some asymptotically flat solution. Fortunately, as was shown in \cite{Konoplya-aether}, the major scattering properties of fields, including the low-lying quasinormal modes, depend on the behavior of the effective potential only in some region near the black hole (if the black hole is not an anti-de Sitter), while the form of the effective potential far from the black hole has no impact on the results. This can easily be explained because the process of scattering occurs mainly near the maximum of the potential barrier.

When computing QNMs with the help of the Frobenius method, we do not take into account this peculiarity of infinity and treat infinity as a mathematical one. Thus, in addition to a number of damped stable modes, we must find some ``unstable'' modes by the Frobenius method. Indeed, we find and tabulate them in Table I, where one can also find estimations of QNMs derived using the WKB method. Unlike the Frobenius method, the WKB formula \cite{WKBorder} implies that one has a positive definite decaying potential at infinity. Therefore, the WKB formula mainly approximates the behavior near the maximum of the potential and does not reproduce those ``unstable'' modes. That is physically adequate as the growing modes appear within Frobenius approach only due to improper ``extension'' of the wave equation (\ref{radialpart}) outside the region of  its validity.

Finally, another instability occurs due to the so-called superradiance: the massive field has a local minimum far from the black hole which works as an effective potential wall, so that the wave amplified due to extraction of rotational energy of the black hole (superradiance) can be reflected back from the distant wall. Repetition of this process leads to unbounded growth of the perturbation. Superradiant instability is shown to be always negligibly small \cite{superradiant}, so that the evaporation time of miniature black holes is much shorter than the characteristic time of the instability growth. For large, astrophysical black holes, superradiant instability of massive fields means that the quantum field will go over to the higher non-superradiant state. In addition, unstable modes are not in the quasinormal sector of the black hole spectrum. Therefore, we do not need to give a detailed analysis of superradiant modes here. Moreover, such an analysis would be technically inaccurate within our approach because the effective potential is known only in some proximity of the black hole and is not exactly known far from the black hole where the local minimum is localized.

Quasinormal modes of the Kerr-Newman black hole immersed in a magnetic field for massive charged scalar field will be determined by a number of parameters, seven altogether: the black hole parameters $Q$, $M$, and $a$, the magnetic field $B$, the scalar field parameters $\mu$ and $e$, and its quantum numbers $m$ and $\ell$. Therefore, complete investigation of the quasinormal modes correlation on these parameters would include an enormous amount of numerical data. We shall show here only the most representative plots for dependence of QNMs on various parameters. We present all our quantities in units of the black hole horizon.

In Figs. \ref{Fig2}, \ref{Fig3}, \ref{Fig4}, \ref{Fig5} one can see for the $Q=0$ case that the $m =0$ modes and the modes with nonvanishing azimuthal number behave quite differently. Actually, modes with $m=0$ (Figs. \ref{Fig2}, \ref{Fig3}) have decreasing damping rate as the angular momentum per unit mass $a$ increases. In the regime of relatively small values of $e B$ the damping rate increases roughly linearly with $e B$. The real oscillation frequency $\Re{\omega}$ decreases linearly with $e B$ and also decreases for growing $a$. We can also see that $e B$ coupling has greater influence on $\Re{\omega}$ than on $\Im{\omega}$, which remains almost unchanged within the region of small $e B$.

For modes with $m > 0$, both $\Re{\omega}$ and $\Im{\omega}$ linearly decrease with $e B$ (Fig. \ref{Fig4}), with one peculiarity: for moderate negative values of $e B$ the $\Re{\omega}$ is not monotonically decreasing with $a$ for all $e B$ anymore. In a large region of both positive and negative values of $e B$, the $\Re{\omega}$ monotonically decreases with $a$ up to some minimal value and then increases. This explains the intersection of curves in Fig. \ref{Fig4}. For negative $m$, we did not observe such a minimum (Fig. \ref{Fig5}) and the behavior is quite similar to the $m=0$ case.

\begin{figure*}
\resizebox{1\linewidth}{!}{\includegraphics*{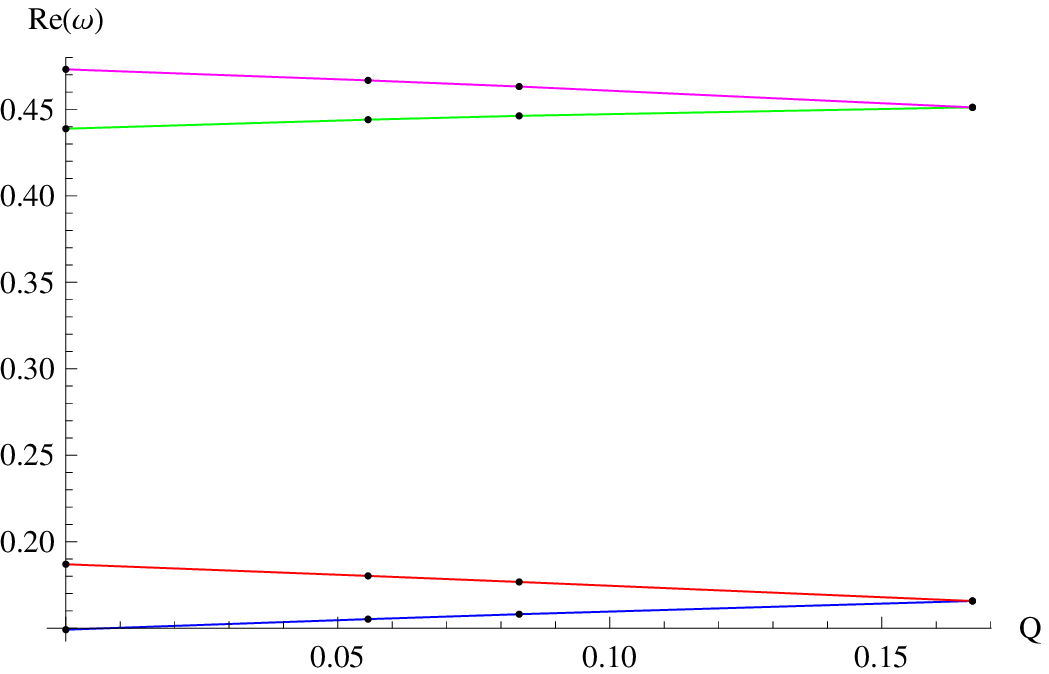}\includegraphics*{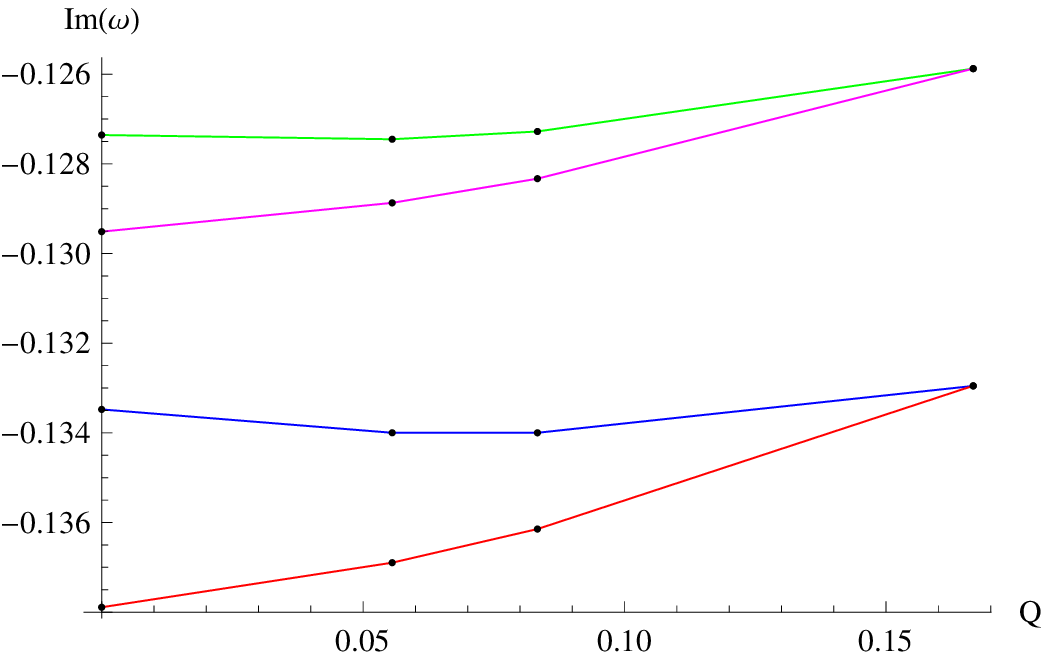}}
\caption{Real and imaginary parts of the fundamental ($n=0$) QNM as a function of $Q$ for $Q=0,2aMB/3,aMB,2aMB$, $B=0.2$, $a=0.6$, $\mu=0$, $e=+0.2$: $\ell=0$ (red, right panel: bottom), $\ell=1$ (magenta, left panel: top) and $e=-0.2$: $\ell=0$ (blue, left panel: bottom), $\ell=1$ (green, right panel: top), $m=0$.}
\label{Fig6-7}
\end{figure*}

\begin{figure*}
\resizebox{1\linewidth}{!}{\includegraphics*{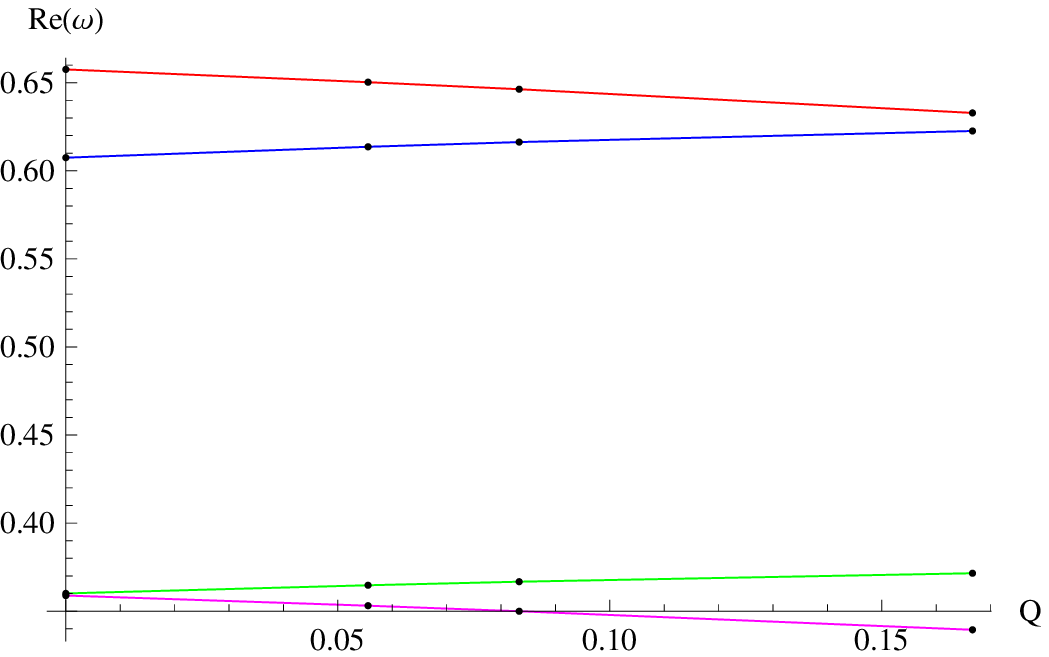}\includegraphics*{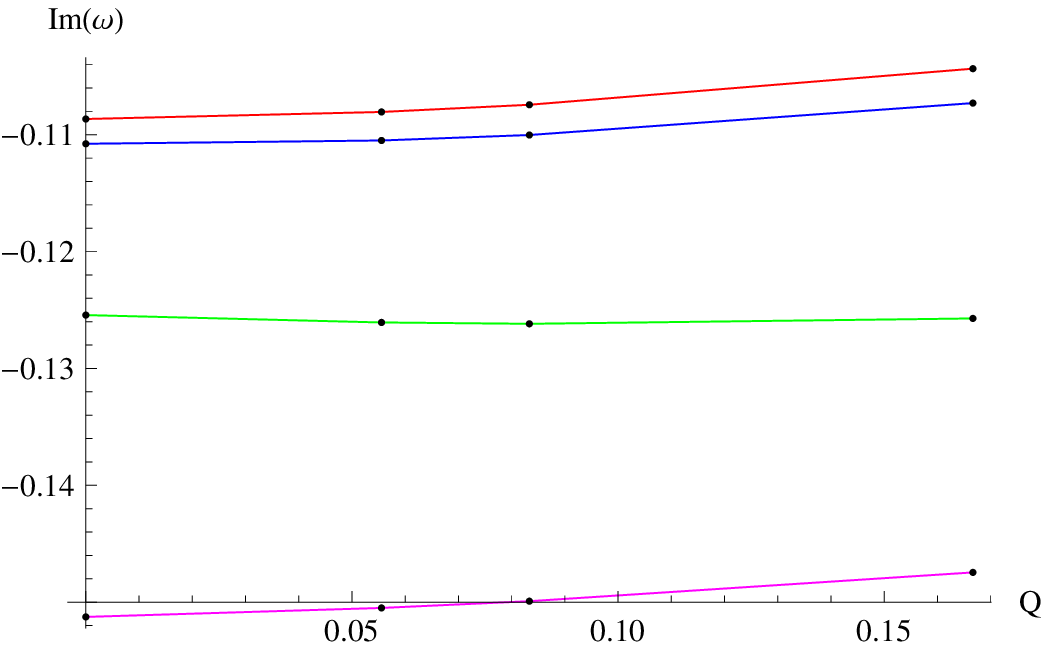}}
\caption{Real (left panel) and imaginary (right panel) parts of the fundamental ($n=0$) QNM as a function of $Q$ for $Q=0,2aMB/3,aMB,2aMB$, $B=0.2$, $a=0.6$, $\mu=0$, from top to bottom: $\ell=m=1$ $e=+0.2$ (red) and $e=-0.2$ (blue), $\ell=-m=1$ $e=-0.2$ (green) and $e=+0.2$ (magenta).}
\label{Fig8-9}
\end{figure*}

\begin{figure}
\resizebox{1\linewidth}{!}{\includegraphics*{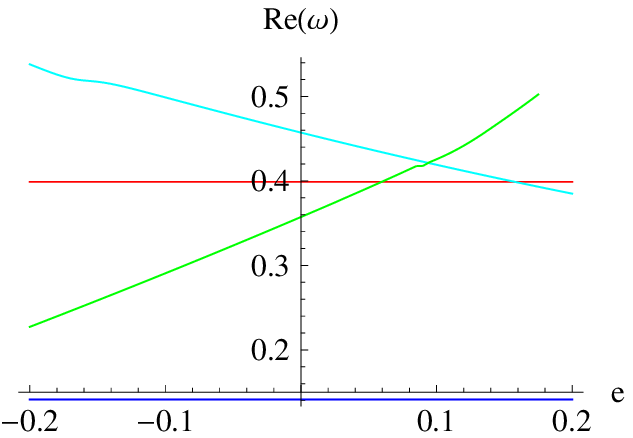}\includegraphics*{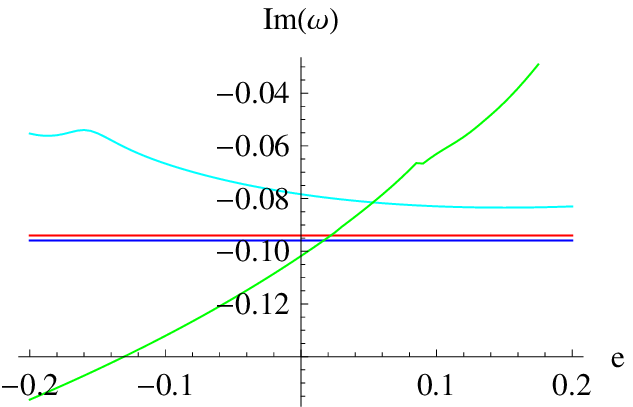}}
\caption{Real (left panel) and imaginary (right panel) parts of the fundamental ($n=0$) QNM as a function of $e$ for $\mu=0.1$, $a=0.25$
$Q=0.9$, $B=1440/749\approx1.92$, $l=m=0$ (blue, bottom horizontal) $l=1,~ m=0$ (red, top horizontal), $l=1,~ m=1$ (cyan, top), $l=1,~ m=-1$ (green).}
\label{Fig10}
\end{figure}

\begin{figure*}
\resizebox{1\linewidth}{!}{\includegraphics*{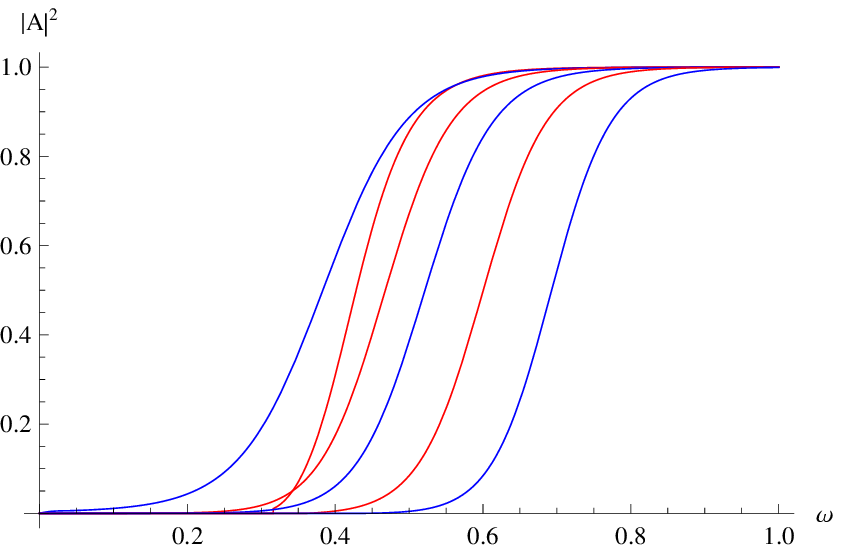}\includegraphics*{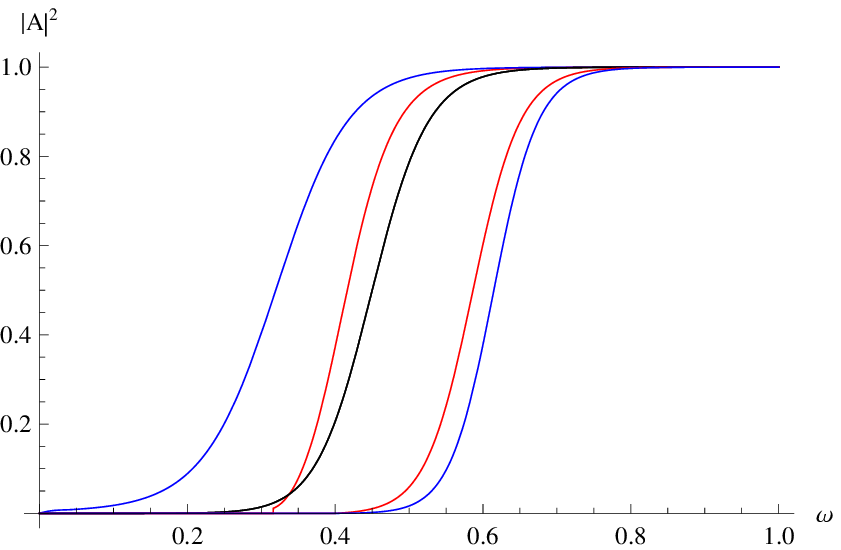}}
\caption{Grey-body factor of the Kerr-Newman black hole (left panel: $a=0.5$, $Q=0.1$, right panel: $a=0.5$, $Q=0.5$) in the magnetic field $B=2/3$ due to massless charged particles ($e=3/20$, $\ell=1$). From left to right: $m=-1$ negative charge particles, $m=-1$ positive charge particles, $m=0$ positive charge particles, $m=0$ negative charge particles, $m=1$ positive charge particles, $m=1$ negative charge particles. For the ``extremal'' black hole $m=0$ grey-body factors for the particles and antiparticles are the same (black line).}
\label{Fig14}
\end{figure*}

In Fig. \ref{Fig6-7}, one can see that $\Re{\omega}$ of the fundamental mode ($\ell = m =0$) is monotonically decreasing with $Q$ for $e B < 0$ and monotonically increasing for $e B > 0$. For $e B < 0$, $\Im{\omega}$ monotonically grows with $Q$, while for $e B > 0$, $\Im{\omega}$ as a function of charge $Q$ decreases up to a minimum at some moderate value of $Q$, and then starts growing. As in the limit $m=0$ and $Q = 2 a M B$ Eqs. (\ref{angularpart}) and (\ref{radialpart}) do not depend on the field charge $e$, the upper and lower curves in Fig. \ref{Fig6-7} coincide for $Q = 2 a M B$. In Fig. \ref{Fig6-7}, the quasinormal behavior for $\ell =1, m=0$ is similar to Fig. \ref{Fig8-9}, where  modes with positive and negative $e$ also coincide in the limit $Q = 2 a M B$. Modes with $\ell = m =1$ have also monotonically decreasing (increasing) $\Re{\omega}$ as a function of the charge $Q$ for $e B < 0$ ($e B > 0$), while $\Im{\omega}$ is monotonically increasing for both positive and negative $e B$ (see Fig. \ref{Fig8-9}).  The same monotonic growth of $\Im{\omega}$ happens for $\ell = 1, m=-1$ mode, so the $\Re{\omega}$ has an opposite behavior: it grows for $e B < 0$ and decreases for $e B > 0$.

Finally, let us discuss the dependence of quasinormal modes on the charge of the field $e$. In Fig. \ref{Fig10}, one can see $\ell = 0, 1$ $m=0$ modes and $\ell =1$, $m=\pm 1$ modes with a charge $Q$ which is equal to the ``extremal'' value $2 a M B$. Modes with $m=0$ naturally form an almost horizontal line because $e$ enters into the wave equation in combination with $m$ or $Q - 2 a M B$. Thus, exactly in the limit  $Q = 2 a M B$, we have a single mode which is independent of $e$. For $\ell =1$, $m=1$, $\Re{\omega}$ decreases as a function of $e$, so this decrease has some small local peaks at larger negative values $e$. In a similar way, $\ell =1$, $m=-1$ modes have both a real and an imaginary part of $\omega$ which almost monotonically increase up to small peaks for moderate values of $|e|$. For sufficiently small values of $|e|$, the dependence of $\omega$ on $e$ is strictly monotonic.

\begin{figure*}
\resizebox{1\linewidth}{!}{\includegraphics*{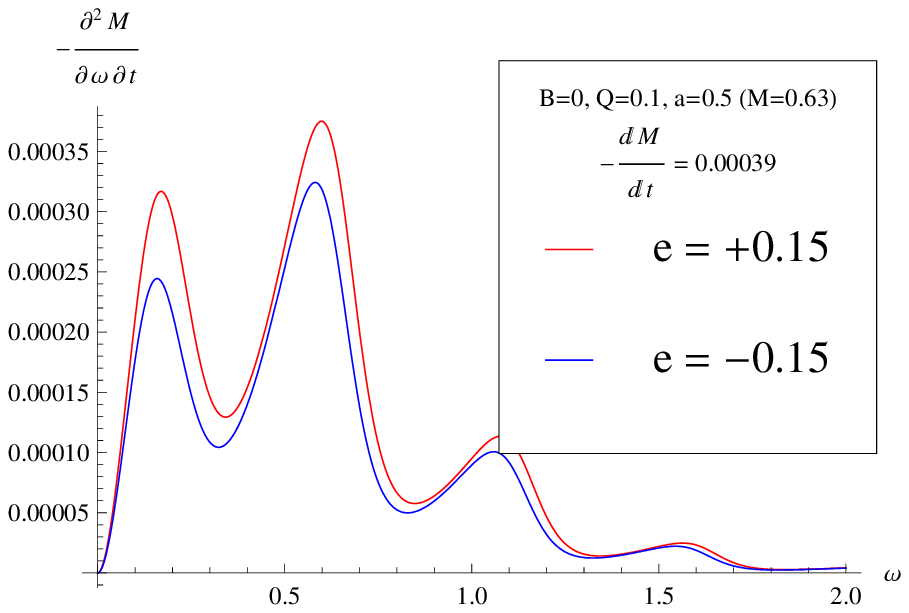}\includegraphics*{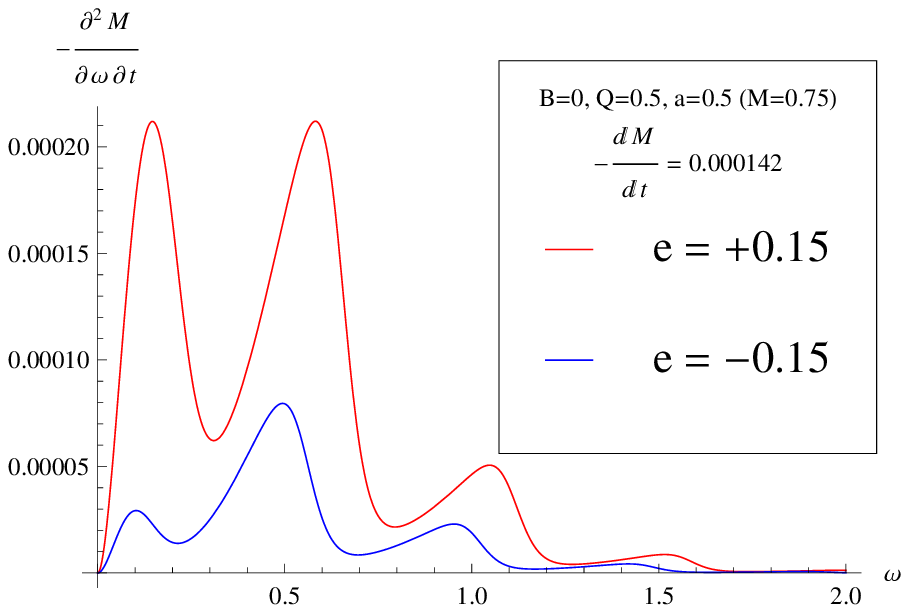}}
\resizebox{1\linewidth}{!}{\includegraphics*{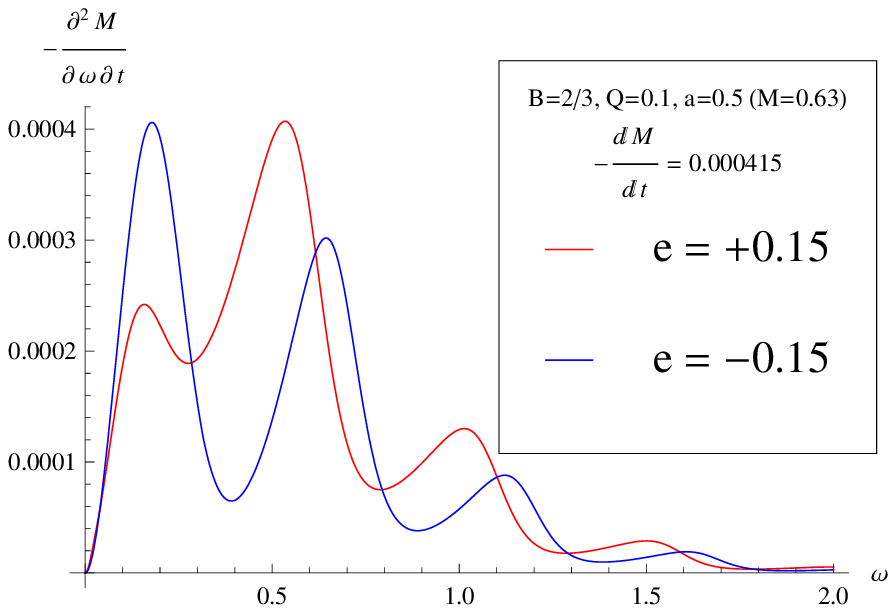}\includegraphics*{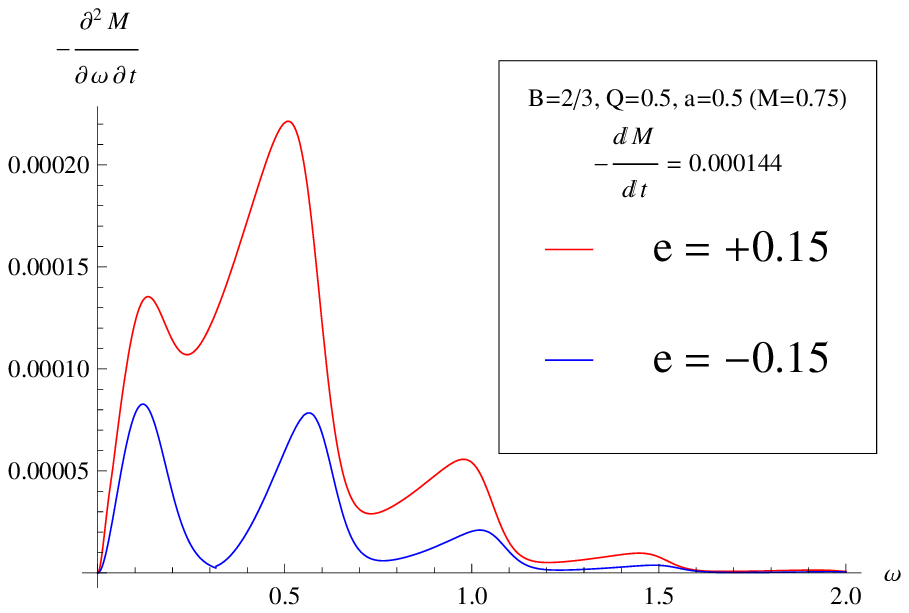}}
\caption{Energy-emission rate of the Kerr-Newman black hole (left panels: $a=0.5$, $Q=0.1$; right panels: $a=0.5$, $Q=0.5$) without the magnetic field (top panels) and with the magnetic field $B=2/3$ (bottom panels) due to massless charged particles ($e=3/20$). The red (top) and blue (bottom) lines correspond respectively to the same and the opposite signs of the charges of the black hole and the emitted particles.}
\label{Fig11}
\end{figure*}

\begin{figure*}
\resizebox{1\linewidth}{!}{\includegraphics*{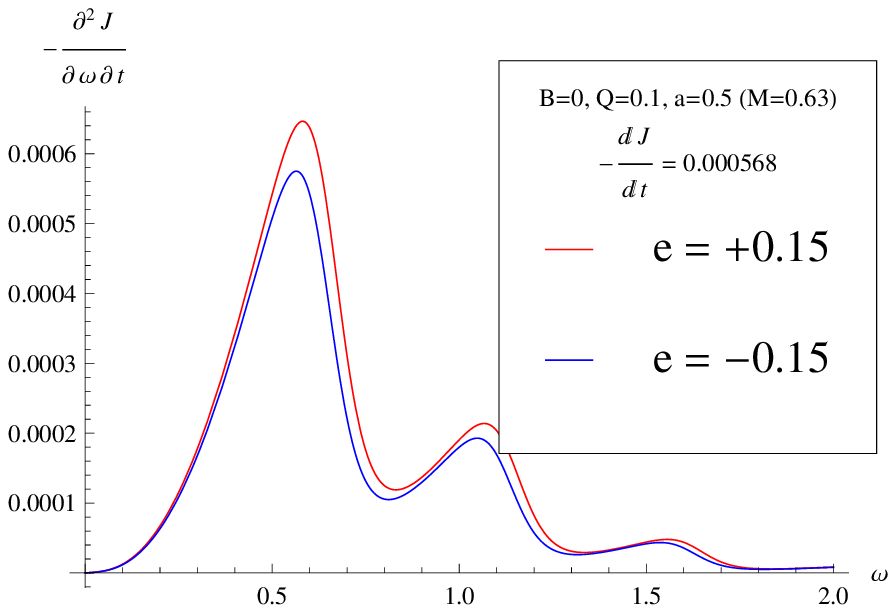}\includegraphics*{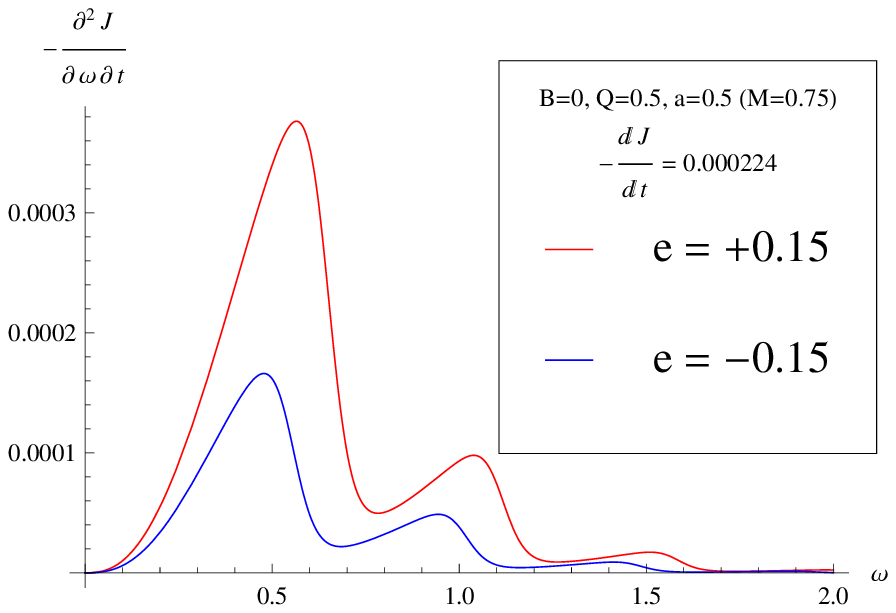}}
\resizebox{1\linewidth}{!}{\includegraphics*{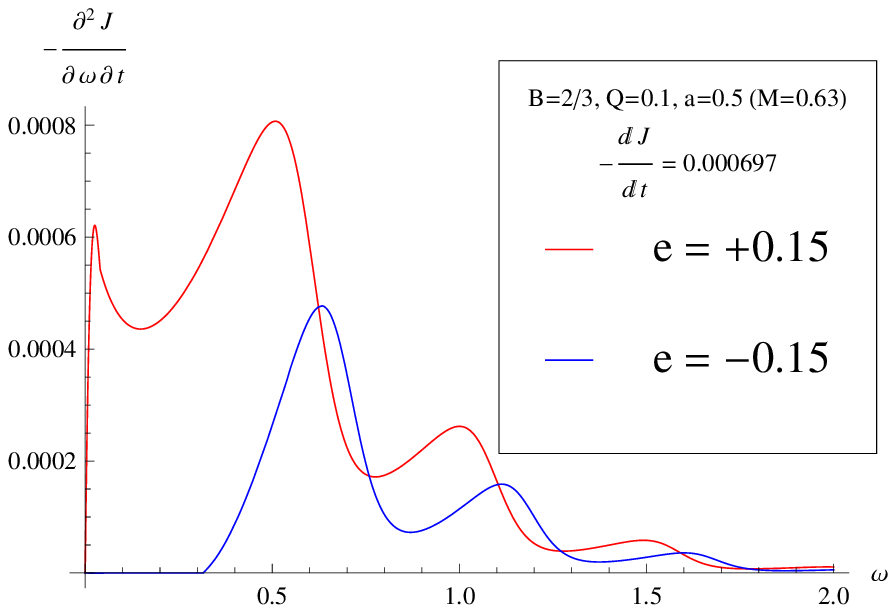}\includegraphics*{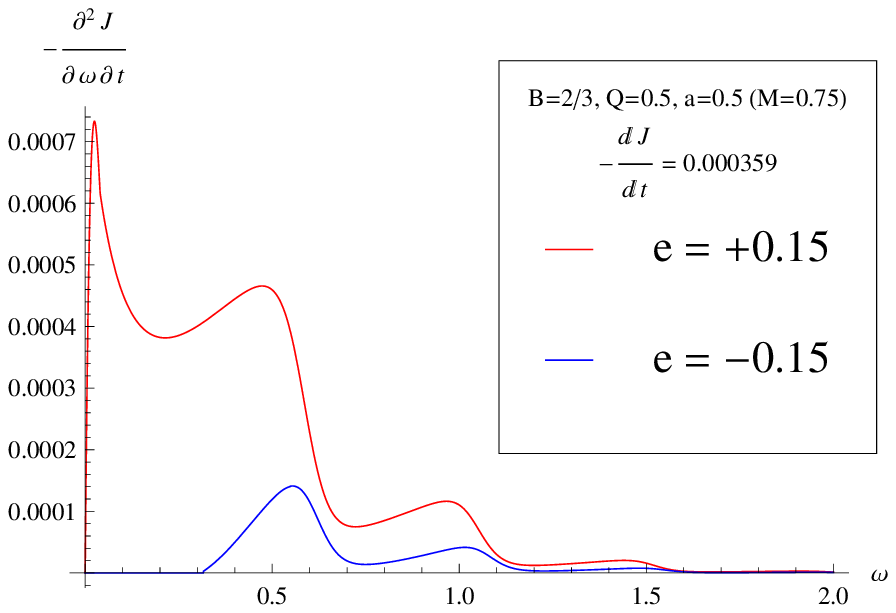}}
\caption{Angular momentum emission rate of the Kerr-Newman black hole (left panels: $a=0.5$, $Q=0.1$; right panels: $a=0.5$, $Q=0.5$) without the magnetic field (top panels) and with the magnetic field $B=2/3$ (bottom panels) due to massless charged particles ($e=3/20$). The red (top) and blue (bottom) lines correspond respectively to the same and the opposite signs of the charges of the black hole and the emitted particles.}
\label{Fig12}
\end{figure*}

\begin{figure*}
\resizebox{1\linewidth}{!}{\includegraphics*{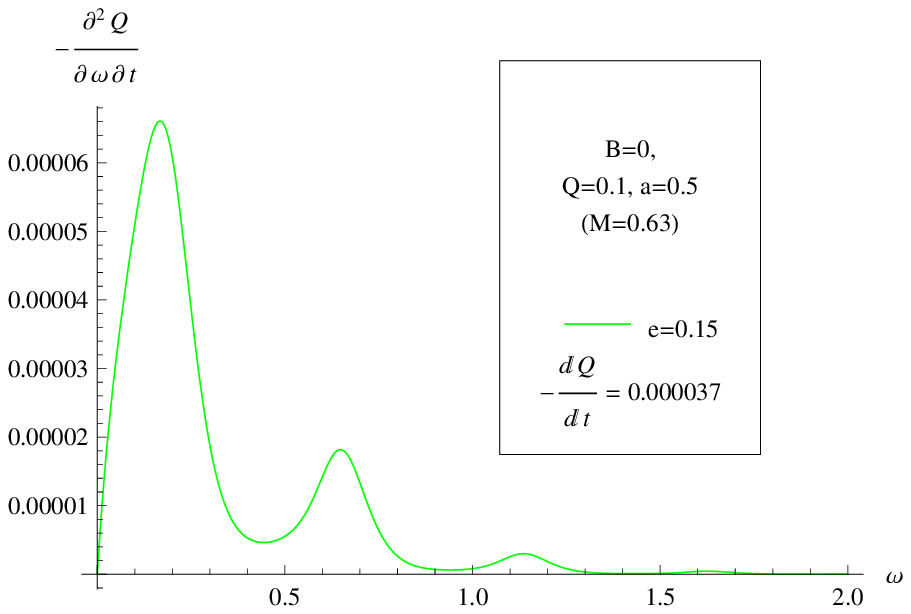}\includegraphics*{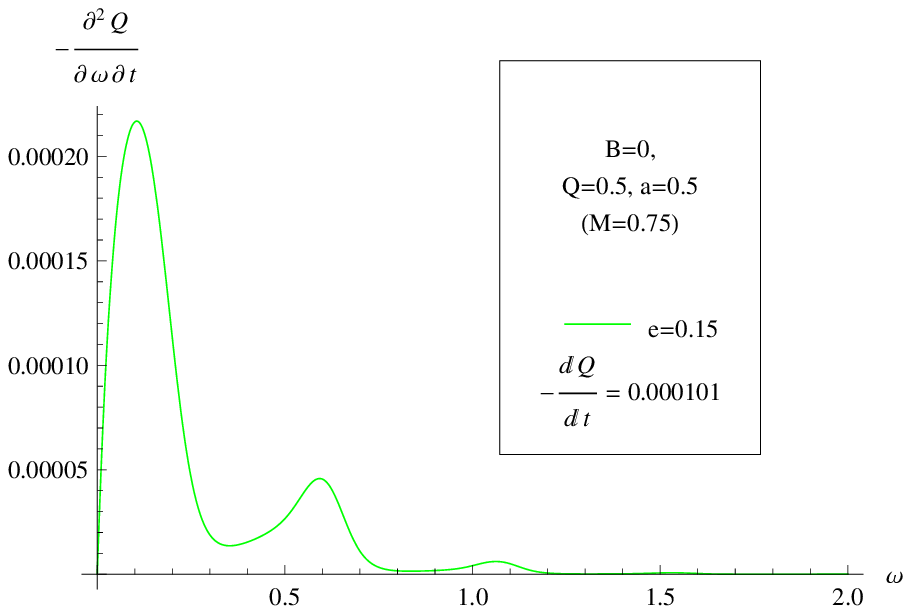}}
\resizebox{1\linewidth}{!}{\includegraphics*{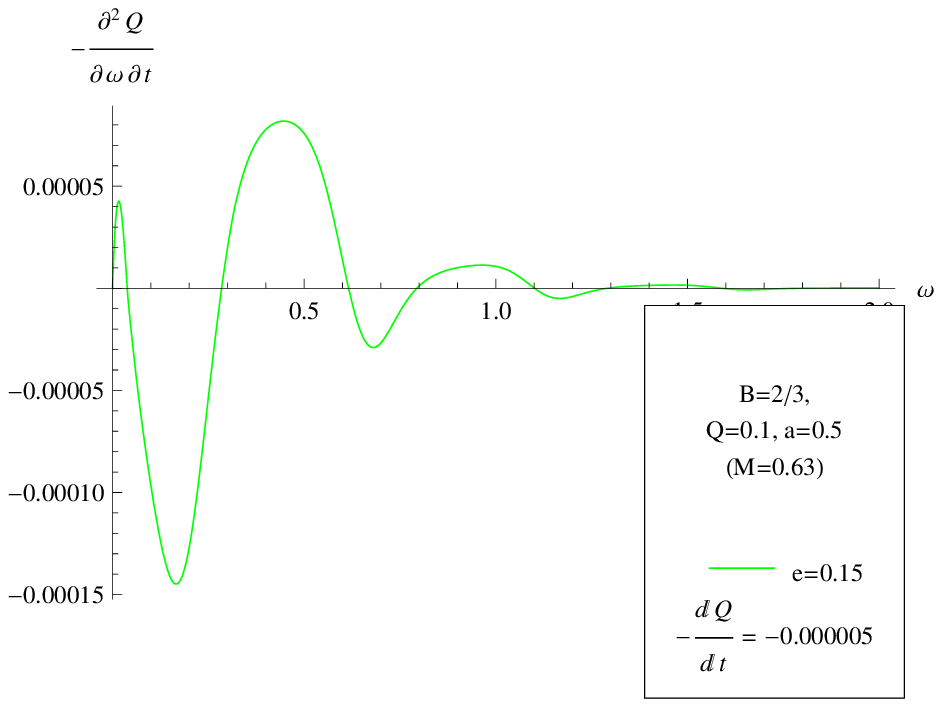}\includegraphics*{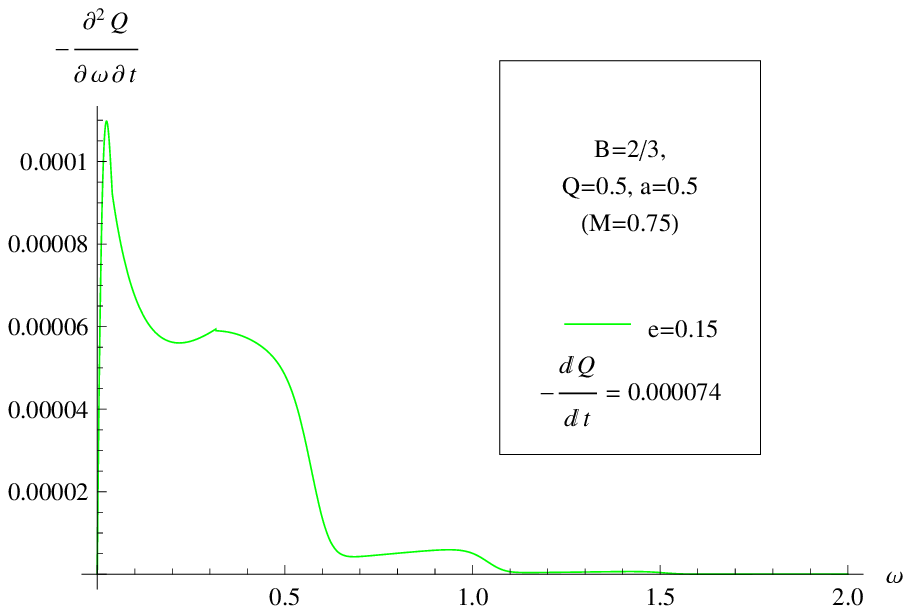}}
\caption{Charge-emission rate of the Kerr-Newman black hole (left panels: $a=0.5$, $Q=0.1$; right panels: $a=0.5$, $Q=0.5$) without the magnetic field (top panels) and with the magnetic field $B=2/3$ (bottom panels) due to massless charged particles ($e=3/20$).}
\label{Fig13}
\end{figure*}

Summarizing, we can say that the Zeeman effect and the Faraday induction influence the quasinormal spectrum in a rather complicated way, where one cannot easily distinguish these two effects. This happens because the magnetic field acts in a different way on modes with different azimuthal numbers $m$: $m=0$ modes are usually only slightly affected by the magnetic field, while $m \neq 0$ are strongly influenced due to the extra coupling term $B m$. Also, the quasinormal behavior is different for negative and positive charge of the field $e$. One common feature of quasinormal modes is that the larger the magnetic field $B$ is, the longer the QNMs live, if $m \leq 0$, while for $m > 0$, they live less long. A similar longer life of the quasinormal modes (yet, for all $m$ and not only for $m \leq 0$) has been observed for the Ernst solution, i.e. for black holes immersed in a strong magnetic field that deforms the geometry \cite{Konoplya:2007yy}. However, in \cite{Konoplya:2007yy} a neutral scalar field was considered, so that the mechanism that influenced the quasinormal spectrum was the induced deformation of the space-time geometry due to magnetic field and not to the Zeeman shift and the Faraday induction. For the Ernst black hole a superradiant instability can also be considerably enhanced \cite{Konoplya:2008hj} for huge values of the magnetic field strength, which seems to be very difficult to achieve. Quasinormal modes of charged scalar fields for nonrotating black holes have also been considered in a number of papers \cite{Konoplya:2002wt,Konoplya:2002ky,Konoplya:2007zx}.

\section{Scattering and Hawking radiation}

A classical black hole in equilibrium does not emit anything. Nevertheless, when considering quantized fields around the black hole, the Hawking radiation appears: a black hole can create pairs of particles from the vacuum on the edge of its horizon. Particles with negative energy go beyond the horizon, while particles with positive energy partially leave the black hole. When analyzing Hawking radiation of black holes we shall assume that the black hole is in thermal equilibrium with its surroundings in the following sense: the black hole temperature does not change between the emission of two consequent particles. This implies the canonical ensemble as a model for the system.

Not all positive energy particles can leave the back hole: part of them is reflected from the potential barrier surrounding the black hole. Thus, the energy-emission rate depends on the grey-body factors which give the fraction of particles penetrating the barrier.

The emission rates for the energy, charge and angular momentum are proportional to the grey-body factors.
The energy-emission rate is
\begin{equation}\label{energy-emission}
-{{dM} \over {dt}} =
\sum_{\ell=0}^\infty \sum_{m=-\ell}^\ell \int{ \left| {\cal A}_{\ell, m} \right|^2 {\omega \over
{\exp (\tilde\omega /T_H ) - 1}}{{d\omega } \over {2\pi }}},
\end{equation}
the charge-emission rate is
\begin{equation}\label{charge-emission}
-{{dQ} \over {dt}} =
\sum_{\ell=0}^\infty \sum_{m=-\ell}^\ell \int{ \left| {\cal A}_{\ell, m} \right|^2 {e \over
{\exp (\tilde\omega /T_H ) - 1}}{{d\omega } \over {2\pi }}}.
\end{equation}
and the angular-momentum emission rate has the form
\begin{equation}\label{momentum-emission}
-{{dJ} \over {dt}} =
\sum_{\ell=0}^\infty \sum_{m=-\ell}^\ell \int{ \left| {\cal A}_{\ell, m} \right|^2 {m \over
{\exp (\tilde\omega /T_H ) - 1}}{{d\omega } \over {2\pi }}}.
\end{equation}
Here, we perform the summation over all the possible values of the quantum numbers $\ell$ and $m$. The grey-body factors are shown in Fig. \ref{Fig14} as functions of $\omega$. There one can see that, for negative $m$, grey-body factors are larger for negatively charged particles than for positively ones, while, for $m \geq 0$, on the contrary, positively charged particles have larger grey-body factors than negatively charged ones. This is indirect influence of the Zeeman term $e B m$ whose contribution depends on the sign of $m$ and $e$. At first glance, negatively (relatively, the black hole charge) charged particles which are emitted radially should have smaller transmission coefficient than positively charged ones: electromagnetic attraction of opposite charges diminishes the transmission of negative particles. For particles which are radiated in all possible direction, this is certainly not so strict and the coupling with the azimuthal number $m$ becomes important.

Let us first discuss Hawking radiation when the magnetic field is absent. In Figs. \ref{Fig11}, \ref{Fig12}, \ref{Fig13}, one can see the emission rates for mass, angular momentum and charge per unit frequency per unit time, and, in the boxes, the results of integration over frequency $\omega$, that is the total emission rates. When $B=0$, the energy, angular momentum and charge emission rates of positively charged particles are larger than those of negatively charged ones for all values of $\omega$, and, consequently, the total emission rates for positive particles are larger as well. When increasing the black hole charge $Q$, the gap between the positive and negative particles emission rates increases.  Electrostatic repulsion of positive particles by the black hole (being proportional to the charge $Q$) enhances the emission of more positive particles. The total energy emission rate decreases as $Q$ is growing, the same being true for the momentum emission rate. The total emission rates include summation of both positive and negative particles, so that the most interesting correlation occurs for the charge emission rate: when $Q$ grows, the charge emission rate, unlike the energy and momentum rates, increases. In general in geometrical units, the black hole looses its mass more quickly than its charge, reaching thereby the extremal Kerr-Newman state.

When one turns on the magnetic field, the picture of Hawking radiation changes drastically. First, at relatively small values of $Q$, the energy and momentum emission rates of  positive particles are no longer than those of negative ones \emph{for all $\omega$}. At some values of $\omega$, the energy and momentum emission rates due to positive particles are smaller than those of negative ones (see Figs. \ref{Fig11}, \ref{Fig12}, \ref{Fig13}). For large values of charge $Q$, the gap between emission rates of positive and negative particles increases, so that the intensity of emission due to positive particles becomes dominant again. What is more important, the presence of the magnetic field considerably increases the energy and momentum emission rates and, at the same time, considerably decreases the charge emission rate, up to changing the sign of the charge emission rate, which means re-charging of the black hole. This means that, in the presence of magnetic fields, the black hole evaporates much quicker and reaches the extremal state in a much shorter period of time. This is quite evident if one notices that the Faraday induction produces an additional (induced) charge $- 2 a M B$ on the surface of the black hole. This charge is opposite to the black hole charge $Q$ and attracts positively charged particles and repulses negatively charged ones. At sufficiently large values of the magnetic field, the absorbtion of positive particles will dominate over the negative ones, which leads to \emph{increasing} instead of decreasing the black hole charge during the evaporation process. This process considerably decreases the time needed by a black hole to reach its extremal state. This may be a relatively small effect for astrophysical black holes but is not negligible for miniature black holes.

Finally, let us recall that, if the induced electric field is as strong as $\mu^2/e$, the electrodynamic Schwinger mechanism of pair creation will occur. Unlike the Hawking radiation which occurs on the edge of the black hole, the Schwinger process will be contributing in the particle production outside the black hole horizon. The Schwinger production will make positive particles move toward the black hole horizon and make negative particles move outwards. Although we have not done any estimates for this process, qualitatively the Schwinger production should probably enhance the recharging of the black hole and, in this way, make the evaporation even quicker.

%\vspace{3mm}
\section{Conclusions}

We have considered the quasinormal modes, classical scattering (through calculations of reflection/transition coefficients) and Hawking radiation of Kerr-Newman black holes immersed in a homogeneous magnetic field. As the simplest model the charged massive scalar field is considered. The equation of motion allows for separation of variables in quite large region surrounding the black hole, but not to spatial asymptotic infinity. We have shown that quasinormal modes and emission rates are influenced by two main effects: the Faraday induction due to rotation in the magnetic field and the Zeeman effect, which is the energy shift of the particle in the magnetic field. The most interesting feature of the dynamics of black holes is in the considerably increased rate of intensity of the Hawking evaporation when one turns on the magnetic field.

This work can be extended in a number of ways. First, one could consider D-dimensional (preferably simply rotating) Myers-Perry black holes immersed in a magnetic field which is localized on the brane. This could provide more realistic estimates for emission rates and quasinormal frequencies for miniature black holes. In addition, for $D>5$ rotating black holes with all different angular momenta the rotation parameter $a$ is not limited anymore. This suggests interesting phenomena for the regime of high rotation because the Faraday induction $2 a B M$ is also not limited. Then, one could calculate the contribution of the Schwinger pair creations in the emission process at very large magnetic field. In addition, a good approach to a more realistic situation would be to consider of the charged massive Dirac field instead of the scalar one. The interaction of spin of a particle with the magnetic field should lead to new phenomena for the Hawking radiation. An analysis of all these questions is one of our nearest future plans \cite{FutureplansKostasus}.

\acknowledgments{The computations were performed with the help of computers at the Institute of Physics at the University
of S\~ao Paulo (IF USP). R. A. K. was supported by the Alexander von Humboldt Foundation, Germany.}

\end{document}